\documentclass[%
 reprint, amsmath,amssymb, aps,]{revtex4-2}

\usepackage{graphicx}
\usepackage{dcolumn}
\usepackage{bm}
\usepackage{leftidx}
\usepackage{txfonts}
\usepackage{color}
\usepackage{enumitem}
\usepackage{floatrow}

\DeclareUnicodeCharacter{2212}{-}
\newcommand{\dd}{\mathnormal{d}}

\newcommand{\old}[1]{}

\begin{document}

\preprint{APS/123-QED}

\title{Dynamics of charged particles and quasi-periodic oscillations in the vicinity of a distorted, deformed compact object embedded in a uniform magnetic field}
\author{Shokoufe Faraji}
 \email{shokoufe.faraji@zarm.uni-bremen.de}
\author{Audrey Trova}%
 \email{audrey.trova@zarm.uni-bremen.de}
\affiliation{%
 University of Bremen, Center of Applied Space Technology and Microgravity (ZARM), 28359 Germany
}%
%

\begin{abstract}
This work presents the dynamic properties of charged test particles influenced by the gravitational and electromagnetic fields. Accordingly in this work, we concentrate on the static and axially symmetric metric containing two quadrupole parameters. One relates to the central object, and another relates to the external distribution of matter. This metric may associate the observable effects to these parameters as dynamical degrees of freedom. The astrophysical motivation for choosing such a field is the possibility to constitute a reasonable model for an actual situation occurring in the objects' vicinity. To test the role of large-scale magnetic fields in accretion processes, we start by analyzing different bound orbits of timelike orbits under the influence of the system's different parameters. This leads to examining their stability concerning radial and/or vertical oscillations. The main focus is to discuss the effect of magnetic field on the oscillation modes' resonant phenomena using different resonant models for disc-oscillation modes. In the present contribution, we further explore the possibility of relating oscillatory frequencies of charged particles to the frequencies of the high-frequency quasi-periodic oscillations observed in the microquasars GRS 1915+105, XTE 1550-564 and GRO 1655-40 via assuming relevance of resonant phenomena of the radial and vertical oscillations.
\end{abstract}

\maketitle

\section{Introduction}

There is no doubt on the role of the magnetic fields in the study of astrophysical systems. In the processes occurring in the vicinity of compact objects, the magnetic fields can manifest their fingerprint in many ways. For example, the local magnetic field in the thin accretion disc is assumed to be the source of viscosity in the accretion process in MRI simulations \cite{1991ApJ...376..214B}. In fact, many black hole candidates are believed to have an accretion disc forming from conducting plasma which their dynamics can produce magnetic fields e.g.  \cite{2013PhRvE..87d3113C,2012PhRvL.108j1101C,2013PhPl...20e2905C}.

An external magnetic field at a large distance in a finite region can be approximated as a uniform magnetic field \cite{1974PhRvD..10.1680W,2015CQGra..32p5009K}. Such a large-scale magnetic field could be initiated during the early phases of the expansion of the Universe \cite{1992ApJ...391L...1R,2001PhR...348..163G,2010Sci...328...73N,2012PhRvD..86l3528J,2013A&ARv..21...62D}. Further, a compact object near the equatorial plane of a magnetar can be approximated to be in a uniform magnetic field if the magnetar is at a distance large enough \cite{2014PhRvD..90d4029K,2016EPJC...76...32S}. In fact, the motion in a gravitational field and in the presence of an external electromagnetic field was explored in a large variety of studies e.g.  \cite{2008CQGra..25i5011K,2010ApJ...722.1240K,2010CQGra..27m5006K} among many more. In this respect, the fundamental frequencies of the motion of test particles play a crucial role in several important astrophysical phenomena. In particular, these frequencies are considered within a great deal of attempts to explain the existence of high-frequency peaks in the Fourier power spectra of X-ray radiation from accreting compact sources and predict lower and upper frequencies of their pairs  $(\nu_U, \nu_L)$.

The so far unexplained rapid variability (so-called high-frequency quasi periodic oscillations - HF QPOs) is usually supposed to originate from the orbital motion in the innermost parts of an accretion disk, since the peaks of high frequencies are close to the orbital frequency of the marginally stable circular orbit representing the inner edge of Keplerian discs. This rapid variability arises across a large scale of mass of the compact sources including both neutron star and black hole low-mass X-ray binaries (LMXBs), as well as active galactic nuclei \cite{2000ARA&A..38..717V,2003ApJ...585..665H,2006csxs.book..157M,2008Natur.455..369G,2013ApJ...776L..10L,2015MNRAS.449..467A,2018ApJ...860L..10S,2020AcASn..61....2Z}.  

Among many models serving to explain QPOs in the past years is the Relativistic Precession Model (RPM). This model assumes that QPOs are produced by a local motion of accreted inhomogeneities like blobs, and relates the twin-peak QPOs to the Keplerian and periastron precession frequency on an orbit located in the inner part of the accretion disc  \cite{1998ApJ...492L..59S,PhysRevLett.82.17}.


In this regard, the properties of the Keplerian and epicyclic frequencies of the orbital motion have been extensively studied in the context of particle motion underlying the presence of a uniform magnetic field in various spacetimes \cite{1981GReGr..13..899A,1986Ap&SS.124..137A,2004PhRvD..69h4022A,2010ApJ...714..748T,2010CQGra..27d5001B,2012ApJ...760..138T,2013arXiv1309.6396G,2014arXiv1411.4811M,2015CQGra..32p5009K,2016EPJC...76...32S,2016EPJC...76..414A,2019PhRvD.100h4038T,2020Galax...8...76A,2020PhyS...95h5008Y} among many others. 


Although, in general, the correlation between these frequencies is qualitatively fitted by the RPM prediction \cite[see][for a detailed discussion and references]{2022arXiv220304787T}, the RPM suffers some theoretical difficulties to explain for example a relatively large observed HF QPO amplitudes which is often observed \cite{2005MNRAS.357.1288B, 2006MNRAS.370.1140B}. During these years, this model modified in many ways. In 2001, the orbital resonance model was introduced. In this model it was supposed that HF QPOs arise from the resonances between oscillation modes of the accreted fluid \cite{2001A&A...374L..19A,2001AcPPB..32.3605K}. 
In fact, several studies within the HF QPOs framework consider fluid motion instead of a test particle motion. In this respect, a different class of models, deals with the collective motion of accreted matter considering normal modes of thin accretion disk oscillations (so-called diskoseismology) and thick disk (torus) oscillations e.g. \cite{2001ApJ...559L..25W,2001PASJ...53....1K,2003MNRAS.344L..37R,2012ApJ...752L..18W,2013A&A...552A..10S,2018MNRAS.474.3967D,2020MNRAS.497.1066M,2020PASJ...72...38K}. In some cases, the QPO frequencies predicted by a given model can still be expressed in the test particle motion for the epicyclic oscillations of slender accretion tori or with reasonable accuracy for the consideration of discoseismic modes. In principle, it is worth mentioning that it is a rather long way from the test particle motion examination to considering the more realistic case of non-slender tori oscillations that has severe impacts on the predicted QPO frequencies e.g. \cite{2016MNRAS.457L..19T,2020A&A...643A..31K}.

Although, in general, the correlation between these frequencies is qualitatively fitted by the RPM prediction, the RPM suffers some theoretical difficulties to explain, for example, the relatively large HF QPO amplitudes that are often observed.

In this work, we explore properties of the Keplerian and epicyclic frequencies of a charged particle
motion in the background of a distorted, deformed compact object with a relatively weak uniform magnetic field that does not affect the spacetime curvature in the vicinity of the compact object. This metric is the generalization of the so-called $\rm q$-metric up to quadrupole moments and aside from the mass of the central object it has two parameters; namely, distortion parameter $\beta$ and deformation parameter $\alpha$, which are not independent of each other. This area of study has been discussed extensively in the literature \cite{1970JMP....11.2580G,doi:10.1063/1.1666501,QUEVEDO198513,PhysRevD.39.2904,Manko_1990,PhysRevD.90.024041}, among many others. We explain briefly about this metric in Section \ref{space1}. 


In the present work, the effects of the electromagnetic field on the stress-energy tensor is neglected. Concerning the observable phenomena, we assume that the HF QPO frequencies can be expressed in terms of the epicyclic frequencies of the charged particle motion. Considering the presence of magnetic field along with a broader spacetime description that can deviate from the Kerr case, we aimed to explore various possibilities of explaining the observed QPOs frequencies with the $3:2$ ratio. 

Furthermore, using the Hamiltonian formalism of the charged particle dynamics, we examine bound orbits via the effective potential of the gravitational field combined with the uniform magnetic field. We are particularly curious about the dynamics regime of motion and its changes with the different combinations of parameters. In addition, we have some freedom in defining these dynamical parameters; however, this work mainly focuses on studying the effect of the magnetisation parameter.

Our survey is of particular interest
for several reasons. First, it is assumed that the Schwarzschild or Kerr metrics describe astrophysical compact objects in the relativistic astrophysical study. However, besides these setups, others can imitate a black hole's properties, such as the electromagnetic signature \cite{PhysRevD.78.024040}. Also, astrophysical observations may not be fitted, in general, within the general theory of relativity by using the Schwarzschild or Kerr metric \cite{2019MNRAS.482...52S, 2002A&A...396L..31A}, like as the mentioned ratio of QPOs. 

In addition, due to their strong gravitational field, considering astrophysical environments, compact objects are not necessarily isolated or possess spherical symmetry. In particular, it seems that the present understanding of astronomical phenomena mostly relies on studies of the stationary and axially symmetric models. It has been shown \cite{2004ApJ...617L..45B} that the possible resonant oscillations can be directly observed when arising in the inner parts of accretion flow around a compact object, even if the source is steady and axisymmetric.

Moreover, we believe our setup allows for the constitution of a set of various reasonable prescriptions for the QPO frequencies, and provide a possibility to study the impact of spacetime and magnetic
field parameters on the predicted QPO frequencies analytically.

The paper’s organization is as follows: Section \ref{space1} presents the background spacetime. The dynamics of charged particles is presented in Section \ref{dynamic}. Section \ref{epicir} discusses epicyclic frequencies and stable circular orbits. In Section \ref{models}, we briefly present QPO models and the related set of prescriptions for the QPO frequencies. In Section \ref{datasec}, we compare the expected QPO frequencies to the observational data. Finally, conclusions are summarized in Section \ref{sum5}.



Throughout this work, we use the signature $(-,+,+,+)$ and geometrized unit system $G=1=c$ (However, for an astrophysical application, we will use SI units). Latin indices run from 1 to 3, while Greek ones take values from $0$ to $3$.


\section{Metric of the distorted deformed compact object} \label{space1}

The first static and axially symmetric solution of Einstein's field equation with arbitrary quadrupole moment is described by Weyl solutions \cite{doi:10.1002/andp.19173591804}. Later, Zipoy and Voorhees \cite{doi:10.1063/1.1705005,PhysRevD.2.2119} introduced an equivalent transformation to $\sigma$-metric (or $\gamma$-metric) that can handle analytically. Then, by introducing a new parameter it is known as $\rm q$-metric \cite{2011IJMPD..20.1779Q}.
In this paper, we choose to work on the generalized version of $\rm q$-metric describing a deformed compact object characterized by quadrupole, while surrounded by a static and axially symmetric external distribution of matter in its vicinity up to the quadrupole \cite{universe8030195}. The metric has this form 

\begin{align}\label{EImetric}
	{\rm d}s^2 &= - \left( \frac{x-1}{x+1} \right)^{(1+{\alpha})} e^{2\hat{\psi}} \dd t^2+ M^2(x^2-1) e^{-2\hat{\psi}} \nonumber\\
	 &\left( \frac{x+1}{x-1} \right)^{(1+{\alpha})}\left[ \left(\frac{x^2-1}{x^2-y^2}\right)^{{\alpha}(2+{\alpha})}e^{2\hat{\gamma}}\right. \nonumber\\
	 &\left. \left( \frac{\dd x^2}{x^2-1}+\frac{\dd y^2}{1-y^2} \right)+(1-y^2) \dd{\phi}^2\right],\
\end{align}
where $t \in (-\infty, +\infty)$, $x \in (1, +\infty)$, $y \in [-1,1]$, and $\phi \in [0, 2\pi)$. The function $\hat{\psi}$ plays the role of gravitational potential, and the function $\hat{\gamma}$ is obtained by an integration of the explicit form of the function $\hat{\psi}$. We consider these functions up to quadrupoles as

\begin{align} \label{1111}
\hat{\psi} & = -\frac{\beta}{2}\left[-3x^2y^2+x^2+y^2-1\right],\\
\hat{\gamma} & = -2x\beta(1+\alpha)(1-y^2)\nonumber\\
  &+\frac{\beta^2}{4}(x^2-1)(1-y^2)(-9x^2y^2+x^2+y^2-1).\
\end{align}
By its construction, this metric is valid locally. In fact, this metric has three parameters: the total mass $M$, deformation parameter $\alpha$, and distortion parameter $\beta$. These two parameters are chosen to be relatively small and connected to the $\rm q$-metric and the surrounding external mass distribution, respectively \footnote{There is a typo in the equation for $\hat{\gamma}$ in our paper related to the study of  unmagnetized case \cite{universe7110447}.}. For vanishing $\beta$, we recover the $\rm q$-metric, and in the case of $\alpha=\beta=0$, the Schwarzschild metric is retrieved. This metric may links the observable effects to the system due to taking these parameters as the new dynamical degrees of freedom. In addition, we try to minimize computational time and numerical errors, since we have some freedom to choose these relatively small dynamical variables. 

The relation between the prolate spheroidal coordinates $(t, x, y, \phi)$, and the Schwarzschild coordinates $(t, r, \theta, \phi)$ reads as

\begin{align}\label{transf1}
 x =\frac{r}{M}-1 \,, \quad  y= \cos\theta.\,
\end{align}
In the rest of the work, we explore this space-time by analyzing the dynamics of particles motion in the presence of an asymptotically uniform magnetic field.


\section{Dynamic of charged particle in a uniform magnetic field}\label{dynamic}

In this work, we consider a weak magnetic field to have no influence on the background space-time. Explicitly, we are interested in the dynamics that occur when a static, axisymmetric central compact object is embedded in an asymptotic uniform magnetic field of strength $B$ aligned with the central body's symmetry axis, and nonsingular throughout the exterior region \cite{PhysRevD.10.1680}. In general, by comparing the compact object's size with the typical length of varying the strength's electric and magnetic fields, we can define a test particle.

Furthermore, the motion of neutral test particles that is not influenced by magnetic fields satisfying

\begin{equation}
    B_G \sim 10^{19}\left(\frac{M_{\odot}}{M}\right) \quad  G.
\end{equation}
This condition comes from comparing the central body's gravitational effect and the strength of the magnetic field $B$ on its vicinity, and for most astrophysical black holes is perfectly satisfied \cite{2011AstBu..66..320P}. In addition, the relative strength of the Lorentz and gravitational forces acting on a charged particle moving in the vicinity of a black hole can be characterized by a dimensionless quantity $b$ which is identified as the
relative Lorenz force by the order of

\begin{equation}\label{b}
    b\sim 4.7 \times 10^7 \left(\frac{q}{e}\right)\left(\frac{m_p}{m}\right)\left(\frac{B}{10^8}\right)\left(\frac{M}{10M_{\odot}}\right),
\end{equation}
where $m_p$ is the mass of a proton, $m$ and $q$ are the mass and charge of the particle. The ratios in this quantity suggest that this characterization is relevant and cannot be neglected for the astrophysical scales.



We start our analysis by stating the standard electric-magnetic tensor as

\begin{equation} \label{emtensor}
F_{\mu \eta} = \partial_{\mu}A_{\eta} -  \partial_{\eta}A_{\mu}.
\end{equation}
As mentioned, following \cite{PhysRevD.10.1680} the external asymptotically homogeneous magnetic field is chosen to be along the polar axis and described by the $\phi$-component of the vector potential. In this metric $ A_{\phi}$ is obtained as

\begin{equation}\label{aphivector}
    A_{\phi}= \frac{1}{2}B (x^2-1) e^{-2\hat{\psi}}
	 \left( \frac{x+1}{x-1} \right)^{(1+{\alpha})}.
\end{equation}
The Lorentz equation that describes the charged test particle motion is given by

\begin{equation}
    m\frac{du^{\mu}}{ds} = q F^{\mu}_{\ \nu}u^{\nu},
\end{equation}
where $u^{\mu}=\frac{d}{ds}$ is the four-velocity of the particle and $s$ is the affine parameter.
In the following part, we use the general Hamiltonian approach to describe the effective potential and dynamics of a charged particle in the vicinity of a distorted, deformed compact object embedded in the external uniform magnetic field.


\subsection{Hamiltonian and effective potential on the equatorial plane}

\begin{figure}
  \includegraphics[width=\hsize]{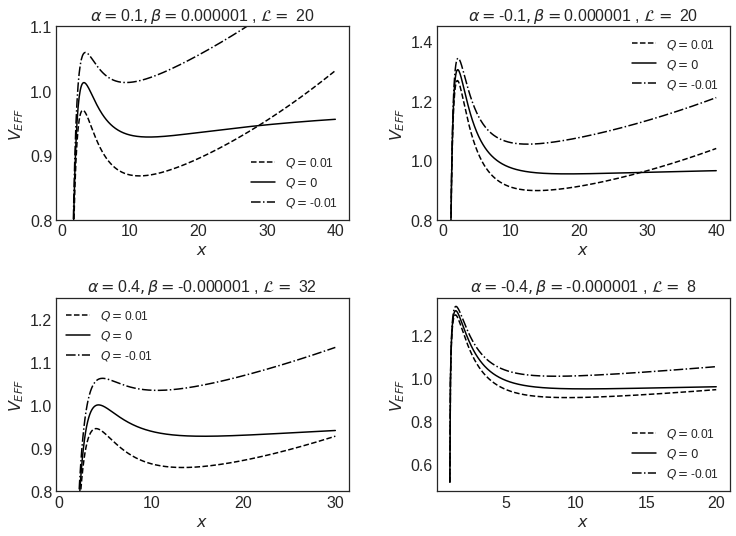}
  \caption{\label{fig1effective} Effective potential for different combinations of the model's parameters: the distortion $\beta$, the deformation $\alpha$ and the magnetic parameter $Q$.}
\end{figure}

The Hamiltonian for the charged particle motion is written as,

\begin{equation}
    H = \frac{1}{2}\left[ (\pi^{\mu}-qA^{\mu})(\pi_{\mu}-qA_{\mu}) + m^2\right],
\end{equation}
where the generalized canonical four-momentum is written in terms of  four-momentum

\begin{equation}
   \pi^{\mu} = p^{\mu} + qA^{\mu}.
\end{equation}
Considering the metric is static and axisymmetric, the conserved quantities are specific energy $E$ and angular momentum $L$ of the particle. They express as follows

\begin{align}
    &E =  -\pi_{t} = \left( \frac{x-1}{x+1} \right)^{(1+{\alpha})} e^{2\hat{\psi}} \frac{dt}{ds} ,\\
    &L = \pi_{\phi} = (x^2-1) e^{-2\hat{\psi}}\left( \frac{x+1}{x-1} \right)^{(1+{\alpha})} \left(\frac{d \phi}{ds}+Q\right) ,\
\end{align}
where $Q:=\frac{qB}{2m}$ is magnetic parameter. In addition, the effective potential  is obtained as

\begin{align}\label{Vei}
    &V_{\rm Eff}=\left( \frac{x-1}{x+1} \right)^{(\alpha+1)} e^{2\hat{\psi}} \left[\epsilon+ \right.\\
    &\left.e^{-2\hat{\psi}} \left( \frac{x+1}{x-1} \right)^\alpha (1-y^2)
    \left(\frac{Le^{2\hat{\psi}}}{(x+1)(1-y^2)} \left( \frac{x-1}{x+1} \right)^\alpha - Q(x+1) \right)^2\right]. \nonumber\ 
\end{align}
The second term corresponds to the central force potential related to $L$, and electromagnetic potential energy related to $B$. In general, we can discuss four different situations in terms of signs of $L$ and $Q$. However, because of the even power in the second term it is sufficient to consider only two situations:

\begin{enumerate}
    \item $LQ>0$, the Lorentz force pushes the particle away in the outward direction with respect to the central object.
   \item $LQ<0$, the Lorentz force pushes the particle in the direction of the z-axis towards the central object.
\end{enumerate}
Moreover, there is a shortcut in the study of the charged particles motion by analyzing the effective potential. In fact, this motion is determined by the energy's boundaries given by $\mathcal{E}= V_{\rm eff}$, where $\mathcal{E}=E^2$. In Figure \ref{fig1effective}, the effective potential presented for different values of parameters $\alpha$ and $\beta$.

In general, different types of orbit dependent on the parameters $\epsilon$, $\mathcal{E}$, $L$, $\alpha$, $\beta$ and $Q$ are possible. In the analytical exploration depending on the number of positive real zeros and the sign of $\mathcal{E}-\epsilon$, one can obtain different types of trajectories.

\begin{figure*}
    \centering
   \includegraphics[width=0.7\hsize]{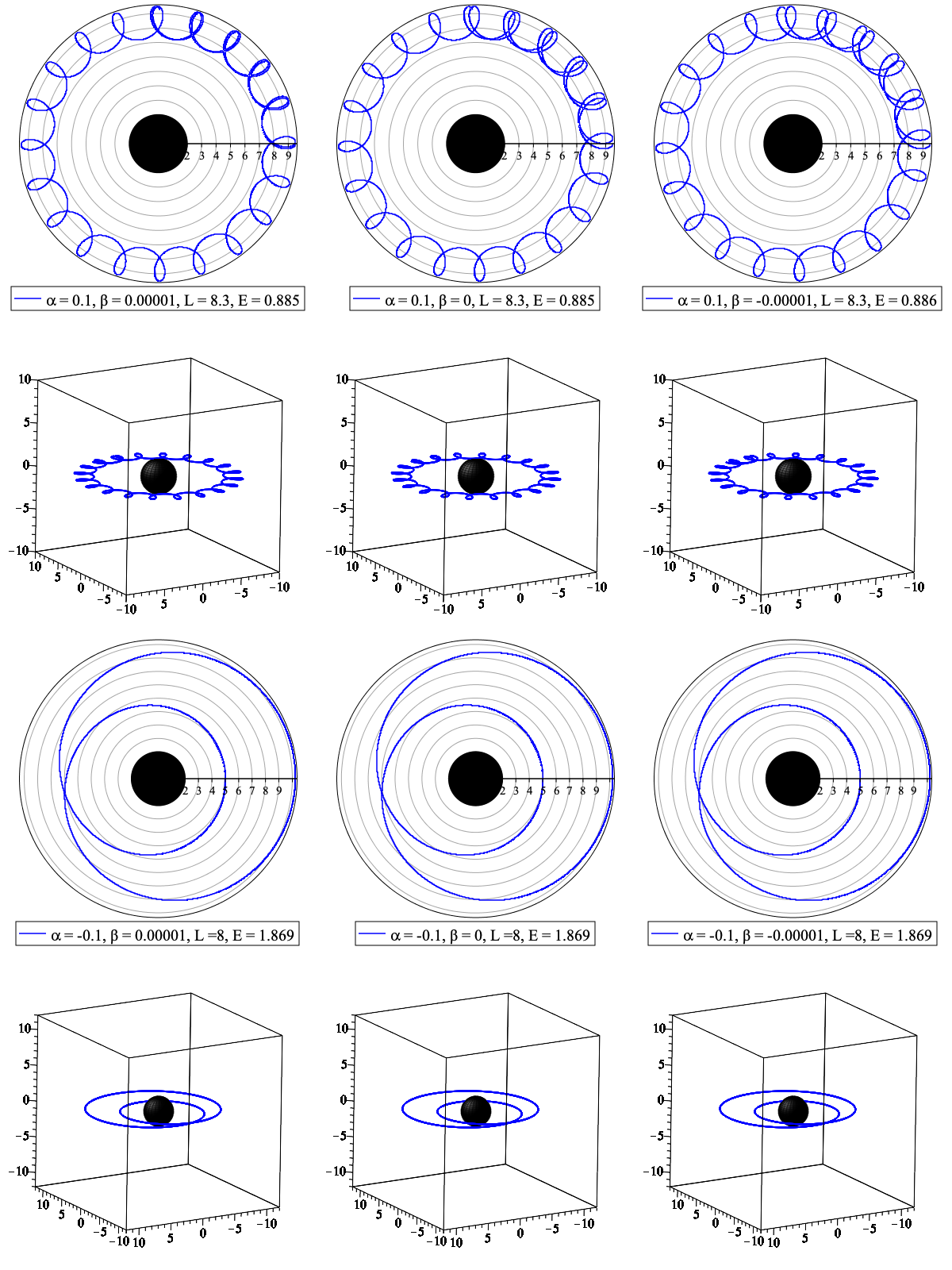}
  \caption{Trajectories of particles for some choices of the parameters. In the first row $Q>0$ and $\alpha>0$, and the initial radius is set to be $r_0=9.5$. In the second one $Q<0$ and $\alpha<0$, and the initial radius is set to be is $r_0=8$.}
    \label{BoundAlphaPosBPos}
\end{figure*}

\begin{figure*}
    \centering
   \includegraphics[width=0.7\hsize]{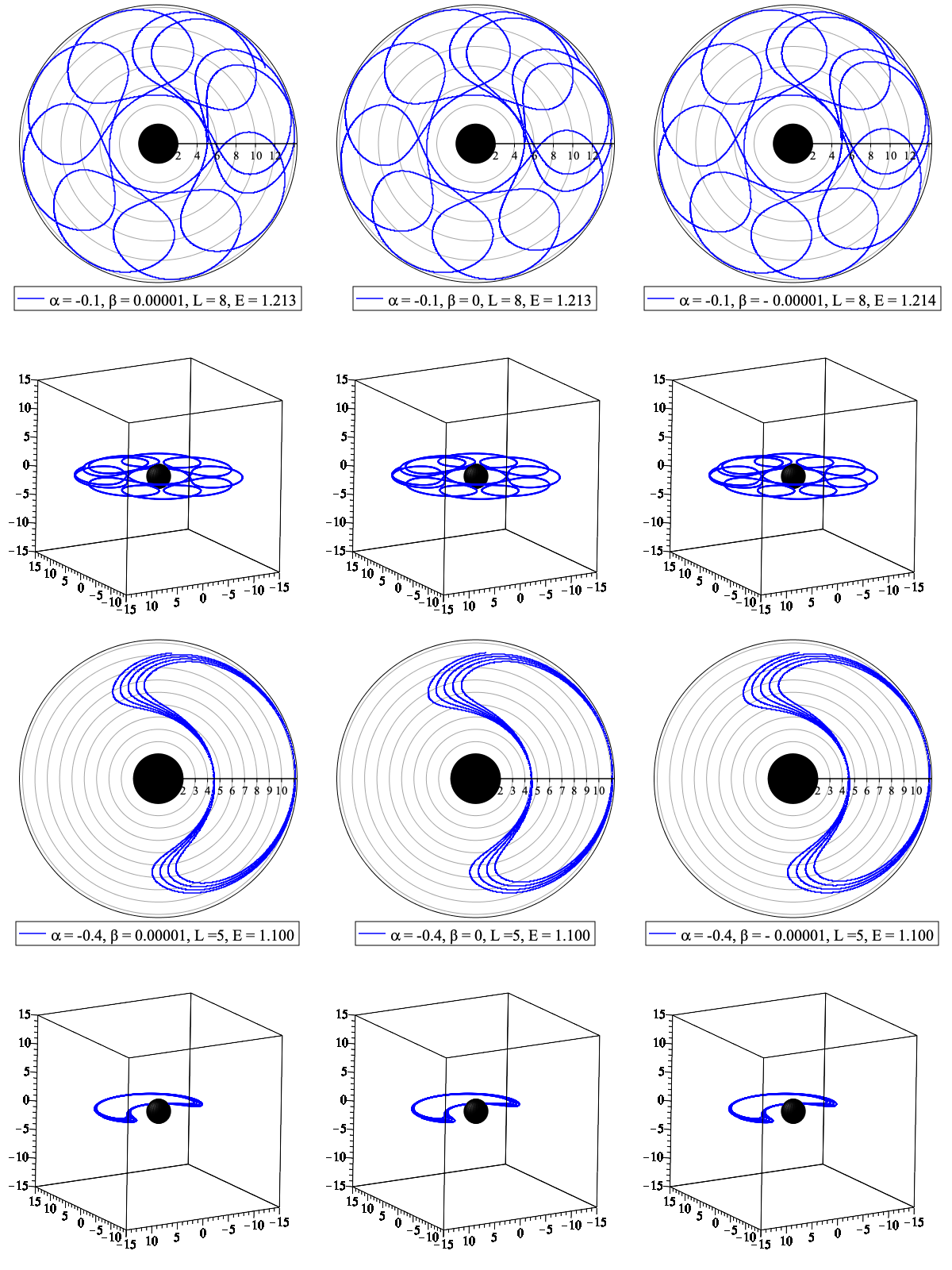}
  \caption{Trajectories of particles for some choices of the parameters. In all plots $Q>0$ and $\alpha<0$. In the first raw, the initial radius is set to be $r_0=5$, and in the second one is $r_0=4.5$.}
    \label{BoundAlphaNegBPos}
\end{figure*}

\begin{figure*}
    \centering
   \includegraphics[width=0.7\hsize]{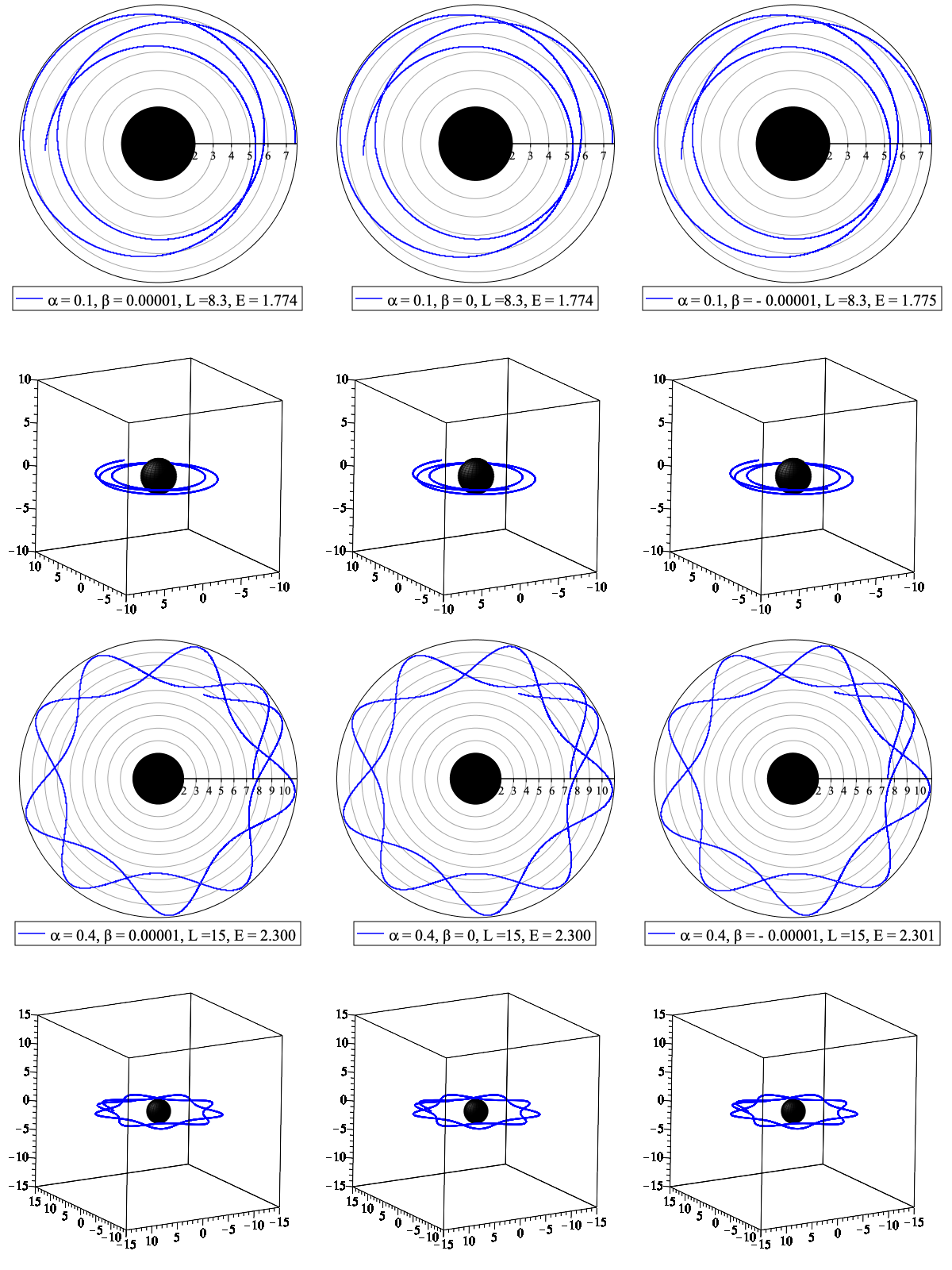}
  \caption{Trajectories of particles for some choices of the parameters. In all plots $Q<0$ and $\alpha>0$. In the plots, the initial radius is set to be $r_0=7.5$.}
    \label{BoundAlphaPosBNeg}
\end{figure*}

We chose to discuss bounded timelike trajectories in this work for their importance in studying the oscillation of particles for a small perturbation in orbit. There is no surprise that test particles' motion can be chaotic in this metric for some combinations of parameters. In fact, when particles are trapped in some region, their trajectories form a toroidal shape around the central object. Figures \ref{BoundAlphaPosBPos}-\ref{BoundAlphaPosBNeg} present some examples of different choices. As we see in the plots, one can obtain different patterns only by a slight change in the value and sign of parameters. In particular, as the Figures \ref{BoundAlphaPosBPos}-\ref{BoundAlphaPosBNeg} suggest, apart from the effect of other parameters, the value for $\alpha$ makes a profound difference in the trajectories. Further analysis reveals that aside from the magnitude, the influence of the signs of magnetic parameter $Q$ along with the sign of $\alpha$ together, have a significant impact on the results.

\section{Epicyclic frequencies and stability of circular motion in a uniform magnetic field} \label{epicir}

There is no doubt on the crucial role of the circular and quasi-circular orbits in the study of accretion processes. In fact, in these processes a variety of oscillatory motions are expected. In this regard, in the study of the relativistic accretion disc three frequencies contribute: The radial frequency $\nu_x = \frac{\omega_x}{2\pi}$, the vertical frequency $\nu_{y} = \frac{\omega_{y}}{2\pi}$, and the Keplerian orbital frequency $\nu_K = \frac{\Omega}{2\pi}$. Furthermore, the resonance among these frequencies may be a source of the chaotic and quasi-periodic variabilities in X-ray fluxes observations.

These fundamental frequencies in the Schwarzschild background, in spheroidal coordinates, are given by
\begin{align} \label{episch}
    &\omega^2_y = \Omega^2= \frac{1}{(x+1)^3} ,\\
 &\omega^2_x = \frac{1}{(x+1)^3}\left(1-\frac{6}{x+1}\right).\
\end{align}
The vertical and keplerian frequencies are positive, but it is not the case for radial frequency. Therefore, the stable circular orbits are located at radial distances larger than the location of the ISCO at $x= 5$ in these coordinates. Besides, for the Schwarzschild solution we have $\omega^2_x<\omega^2_y=\Omega^2$. In fact, in contrast to Newtonian geometry, bounded quasi-elliptic orbits can also exist. Namely, the trajectories relevant to a perturbed stable circular test particle may not be elliptic but exhibit a periapsis shift. This is often called the effect of relativistic precession \cite{2001ASPC..234..213S}.




In what follows, we investigate the stability of circular motion in the presence of the asymptotically homogeneous magnetic field. In this regard, seminal papers \citep{1981GReGr..13..899A,1986Ap&SS.124..137A} studied the existence and stability of circular non-geodesic equatorial orbits, and derived Keplerian and epicyclic orbital frequencies in Kerr(-Newman) geometry. One of the particular models they considered was a rotating black hole immersed in an asymptotic uniform external magnetic field. In this work, following their method we extend the result to this space-time containing quadrupoles. Of course, if parameters of the metric vanish, the axial and radial oscillations frequencies are reduced to the ones given in their works for non-rotating case.

The equation of motion for a particle with mass $m$ and electric charge $e$ is the geodesic equation with force in its right-hand side reads as

\begin{equation}\label{geo}
    \frac{d^2x^{\mu}}{ds^2}+\Gamma^{\mu}_{\nu\rho} \frac{dx^{\nu}}{ds} \frac{dx^{\rho}}{ds}= \frac{q}{m}F^{\mu}_{\ \eta}\frac{dx^{\eta}}{ds}
\end{equation}
In the equatorial plane we have $x=x_0$ and $y=0$, and further we replaced all necessary Christoffel symbols on the left-hand side. For substitute the right hand side we use \eqref{emtensor} and \eqref{aphivector}.

To derive a more general class of orbits in the equatorial plane deviating from the circular ones, one can use the perturbation approach and consider a slightly perturbed orbit $x^{\prime \mu}=x^{\mu}+\xi^{\mu}$ from the original one $x^{\mu}$. By substituting this relation into equation \eqref{geo} and consider only linear orders in $\xi^{\mu}$, we obtain \citep{1986Ap&SS.124..137A}


\begin{equation}
  \frac{d^2\xi^{\mu}}{dt^2}+2\gamma^{\mu}_{\ \eta}\frac{d\xi^{\eta}}{dt}+\xi^{\eta}\partial_{\eta}U^{\mu}= \frac{q}{m u^0}f^{\mu} \label{geocom},
  \end{equation}
where, 
   
 \begin{align}
  &\gamma^{\mu}_{\ \eta} = \left[2\Gamma ^{\mu}_{\ \eta \delta}u^{\delta}(u^{0})^{-1}-\frac{q}{m u^0}F^{\mu}_{\ \eta}\right]_{y=0} ,\\
  &U^{\mu}=\left[\gamma^{\mu}_{\eta}u^{\eta}(u^0)^{-1}-\frac{q}{m u^0}F^{\mu}_{\ \eta}u^{\eta}(u^0)^{-1}\right]_{y=0} ,\
\end{align}
where the external force presents as $f^{\mu}_{\ \nu}u^{\nu}(u^0)^{-1}$, and the four-velocity for the circular orbits in the equatorial plane is $u^{\mu}=u^0(1, 0, 0, \Omega)$ \citep{1986Ap&SS.124..137A}. The integration of equation \eqref{geocom} for the $t$ and $\phi$ components gives us

\begin{align}
    &\frac{d\xi^{\eta}}{dt}+\gamma^{\eta}_{\ \nu}\xi^{\nu} = \frac{q}{m u^0} \int {f^{\eta} dt},\\
    &\frac{d^2\xi^{x}}{dt^2}+\omega^2_x \xi^{x} = \frac{q}{m u^0} \left(f^x - \gamma^x_{\ \eta} \int f^{\eta} dt\right) \label{omegay},\\
    &\frac{d^2\xi^{y}}{dt^2}+\omega^2_x \xi^{y} = \frac{q}{m u^0} f^y .\
\end{align}
where here $\eta$ can be taken as $t$ or $\phi$, and

\begin{align} \label{sqf}
   &\omega^2_{x} = \partial_x U^x - \gamma ^{x}_{\ \eta}\gamma ^{\eta}_{\ x} ,\nonumber\\
   & \omega^2_{y} = \partial_y U^y .\
\end{align}
This system of equations describes radial phase and vertical oscillations of the charged particle around the circular orbits \footnote{For an alternative definition of the epicyclic harmonic motion see \citep{1984ucp..book.....W}.}. The positive sign of the squared frequencies \eqref{sqf} determines the stability of circular orbits; otherwise, even a minimal perturbation can deviate substantially from the unperturbed orbit. In the absence of the external force, these equations describe the free radial phase and vertical oscillations of particles around the circular orbits. In what follows, we analyze the behavior of these frequencies \eqref{sqf}.

\subsection{Properties of epicyclic frequencies \label{sec:PropFreq}}

The correspondence frequencies of the equation \eqref{sqf} in the background of a distorted, deformed compact object, explicitly are written in the appendix \ref{app1}. In this case, as equations \eqref{A1} and \eqref{A2} suggest, we can not write $\omega_x$ and $\omega_y$ as some coefficient of $\Omega$, like the standard procedure in Schwarzschild and Kerr space-times. However, this procedure is possible in this metric only when there is no magnetic field presented. 

\begin{figure}
   \centering
    \includegraphics[width=\hsize]{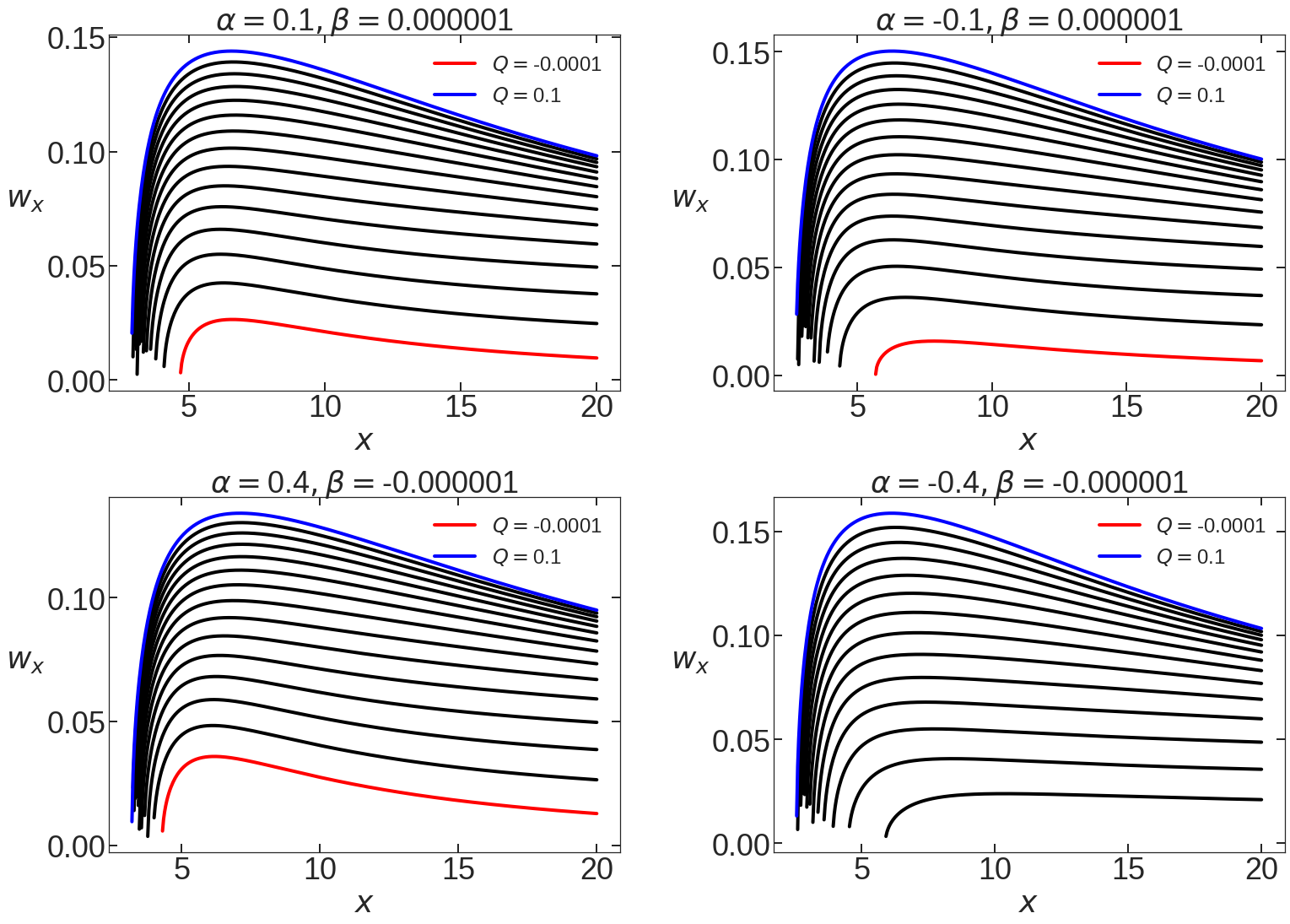}
   \caption{The radial epicyclic frequency $w_x$ is plotted with respect to $x$ for four different pairs of $(\alpha,\beta)$. On each plot, different values of $Q$ are used from $Q=-0.0001$ to $Q=0.1$. The blue curve is corresponds to $Q=0.1$ and the red curve to $Q=-0.0001$. We can note for the pair $(-0.4,-0.000001)$, the red curve $(Q=-0.0001)$ is not appearing. This means that for these values of $\alpha$ and $\beta$, the radial epicyclic frequency is not real.}
    \label{wxext}
\end{figure}

\begin{figure}
   \centering
     \includegraphics[width=\hsize]{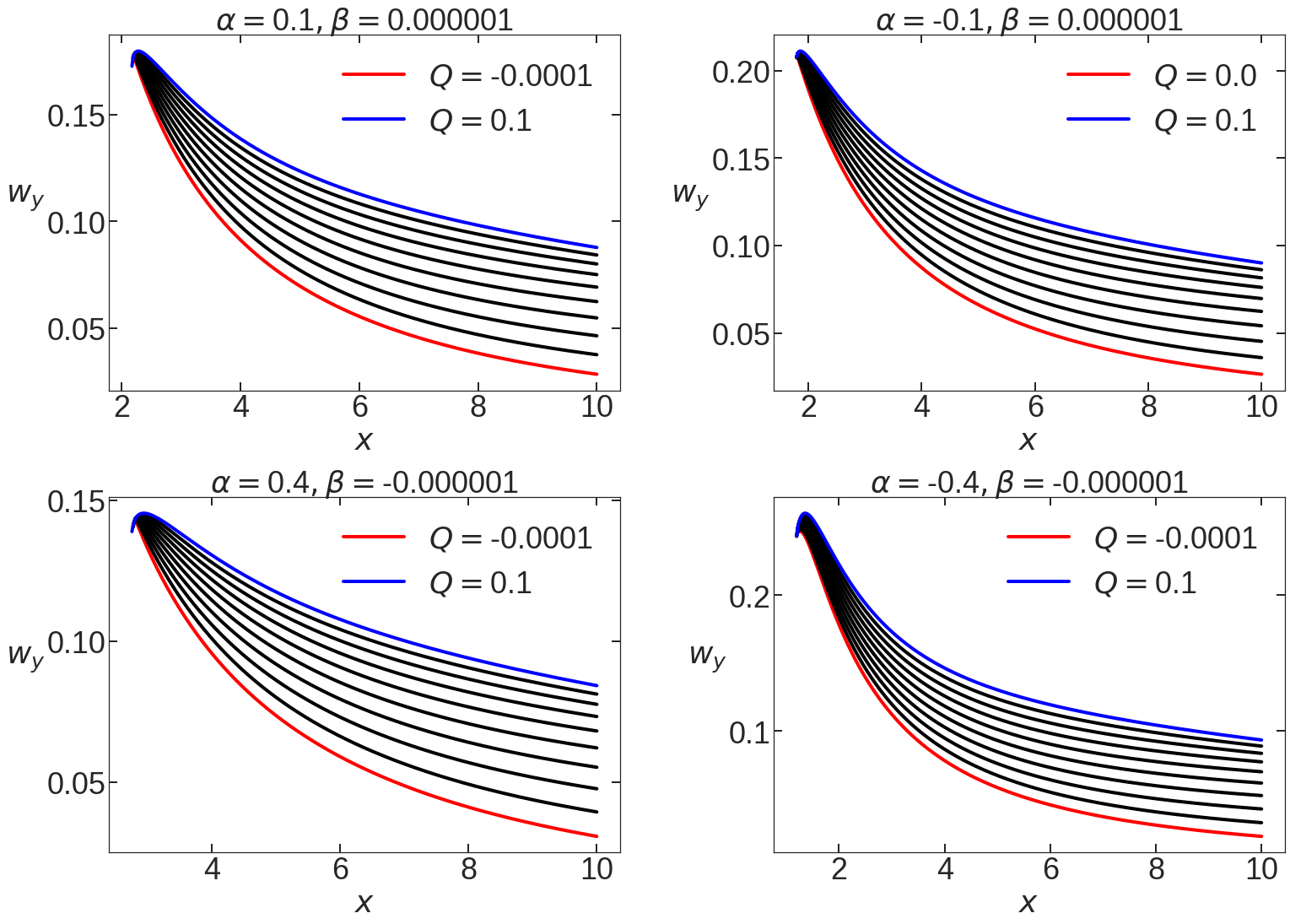}
   \caption{The vertical epicyclic frequency $w_y$ is plotted with respect to $x$ for four different pairs of $(\alpha,\beta)$. On each plot, different values of $Q$ are used from $Q=-0.0001$ to $Q=0.1$. The blue curve corresponds to $Q=0.1$ and the red curve to $Q=-0.0001$.}
    \label{wyext}
\end{figure}

In Figure \ref{wxext} and \ref{wyext} we see the locations of maxima of these frequencies $\omega_x$ and $\omega_y$. For any choice of parameters, the radial frequency's extrema must be located above the marginally stable orbit. The extrema are dependent on the choice of parameters for the vertical frequency, but these are always a monotonic function of distance $x$.

\begin{figure*}
    \centering
    \includegraphics[width=\hsize]{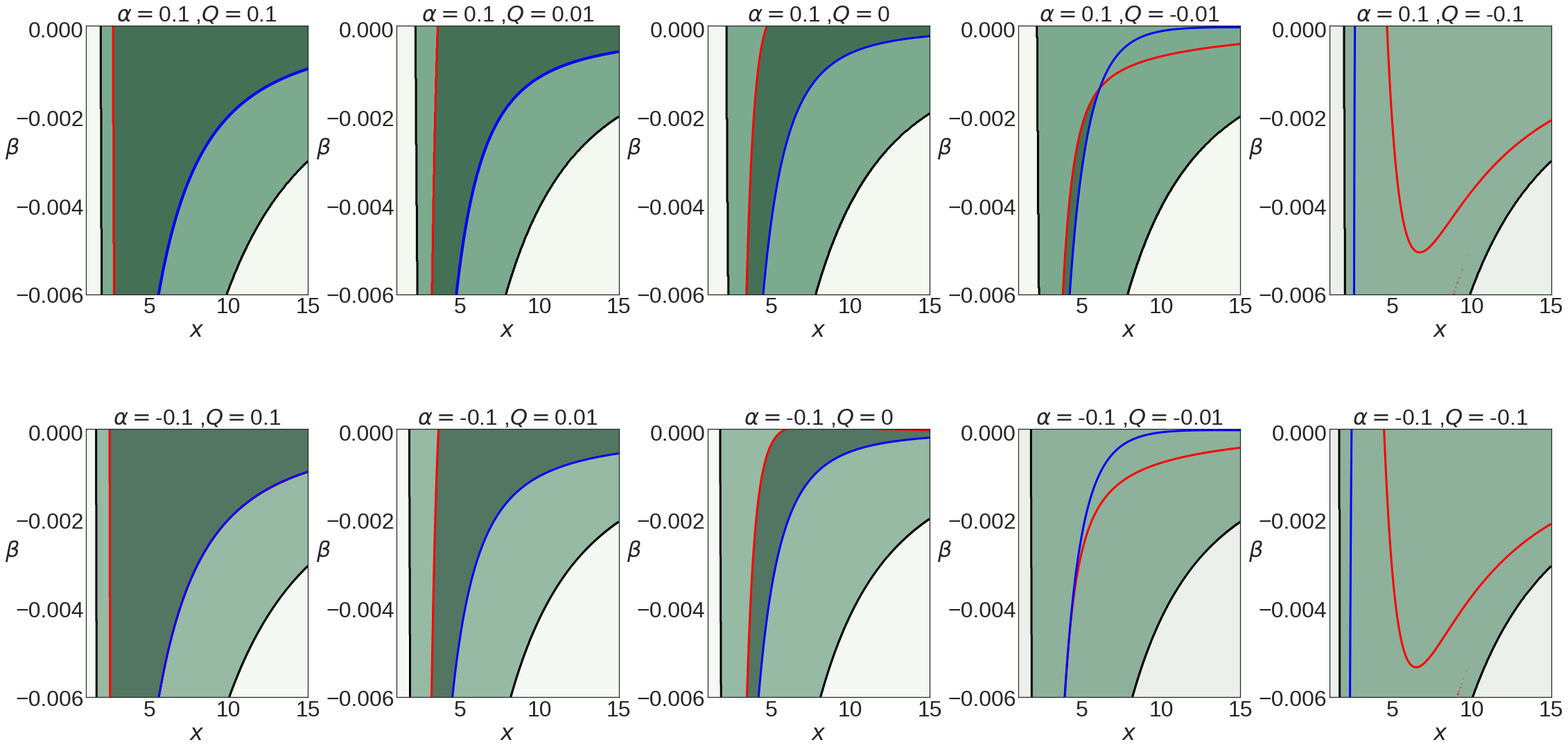}
    \caption{Stability of the timelike circular orbits of a charged particle in the ($x,\beta$)-plane. Timelike circular orbits can exist in the green light area. The blue curve represents $w_y^2=0$ and the red one $w_x^2=0$. Timelike orbits are stable with respect to vertical and radial perturbations in the region where $w_x^2>0$ and $w_y^2>0$. This area is depicted by the green-gray region, which combined the conditions of existence and the condition of stability, $w_x^2>0$ and $w_y^2>0$. The analysis is done for two positive values of $Q$, two negatives values of $Q$, and the unmagnetized cased $Q=0$.}
    \label{figstab1}
\end{figure*}

\begin{figure*}
    \centering
    \includegraphics[width=\hsize]{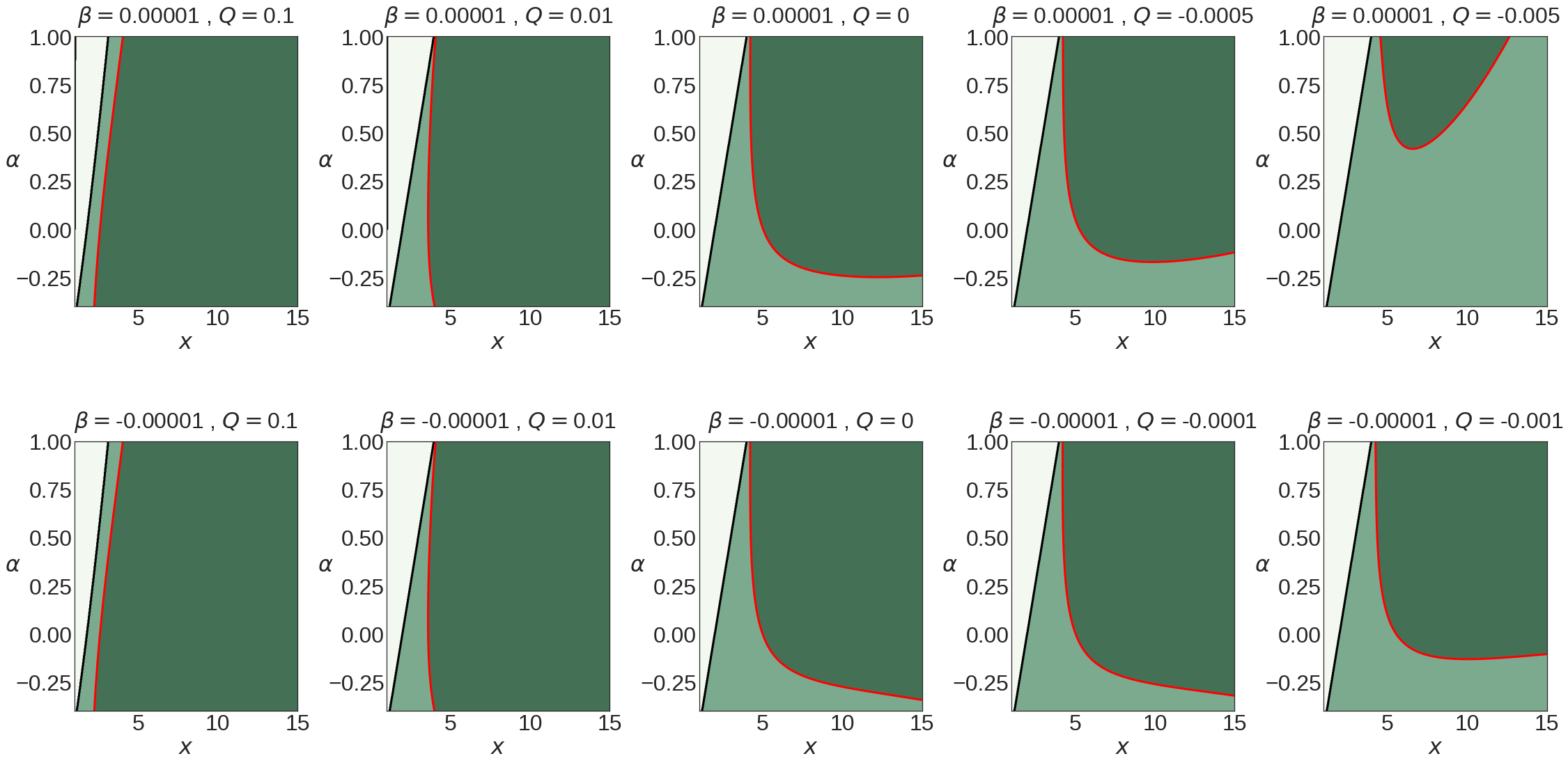}
    \caption{Stability of the timelike circular orbits in the ($x,\beta$)-plane. Timelike circular orbits exist in the green light area. The blue curve represents $w_y^2=0$ and the red one $w_x^2=0$. Timelike orbits are stable with respect to vertical and radial perturbations in the region where $w_x^2>0$ and $w_y^2>0$. The analysis has done for two positive values of $Q$, two negatives values of $Q$, also for the unmagnetized cased $Q=0$.}
    \label{figstab2}
\end{figure*}

Figures \ref{figstab1} and \ref{figstab2} are depicted for the study of the existence and the stability of the timelike orbits with respect to vertical or radial oscillations. All the plots in Figure \ref{figstab1} are in the $(x,\beta)-$plane for two chosen values of the deformation parameter $\alpha$ -one positive and one negative- and for different values of the magnetic parameter $Q$. While Figure \ref{figstab2} represents the region of existence and stability in the $(x,\alpha)$-plane for two chosen values of the distortion parameter $\beta$, and for the same values for the magnetic parameter $Q$ as in Figure \ref{figstab1} .

\begin{figure}
    \centering
    \includegraphics[width=\hsize]{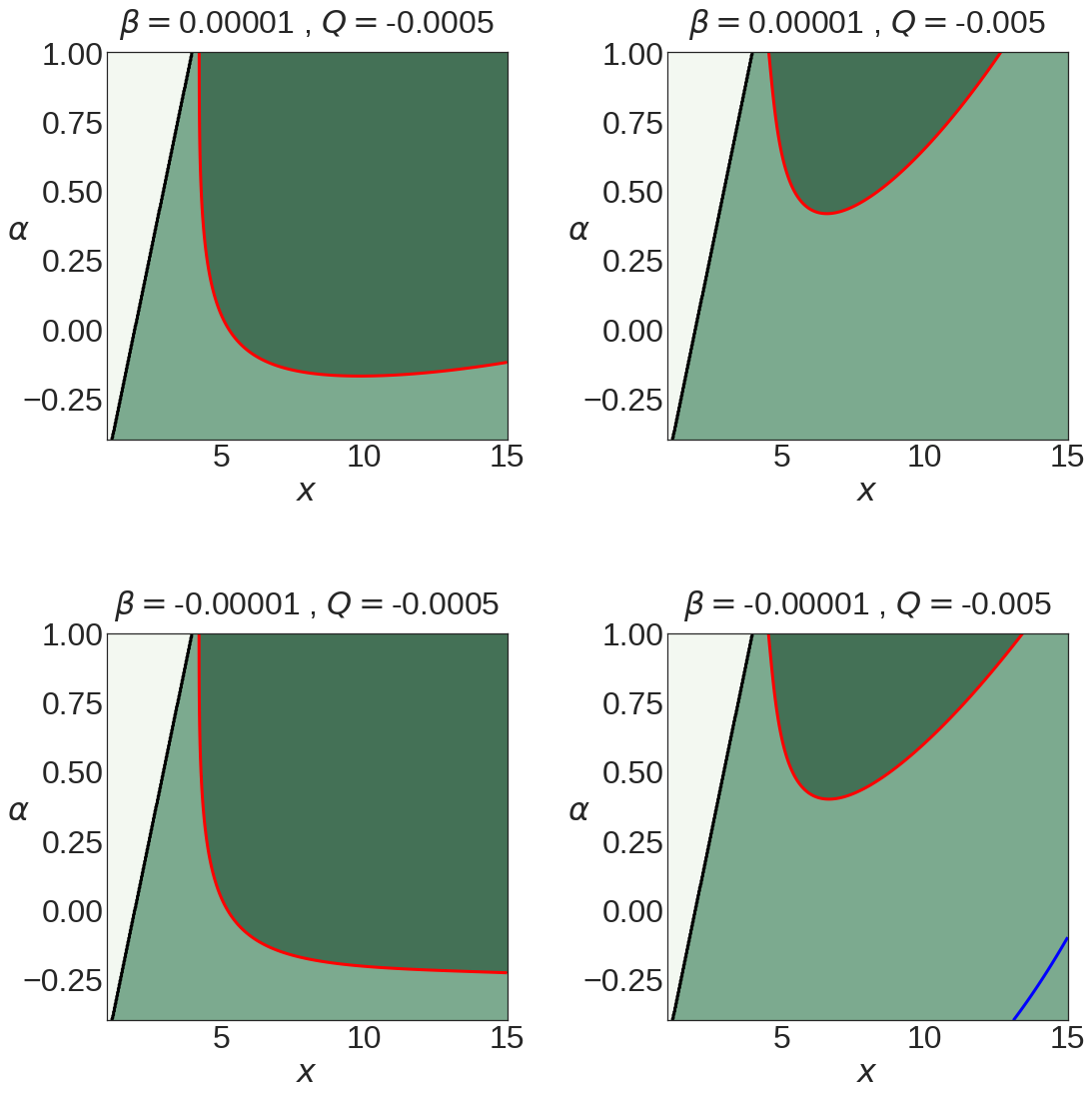}
    \caption{Stability of the timelike circular orbits in the ($x,\alpha$)-plane. We zoomed in for the small negative values of $Q$. These two columns can be inserted between the 3rd and 4th columns of Figure \ref{figstab2}. The description of the regions and curves are the same as Figure \ref{figstab2}.}
    \label{figstab2Zoom}
\end{figure}

\begin{figure*}
    \centering
    \includegraphics[width=\hsize]{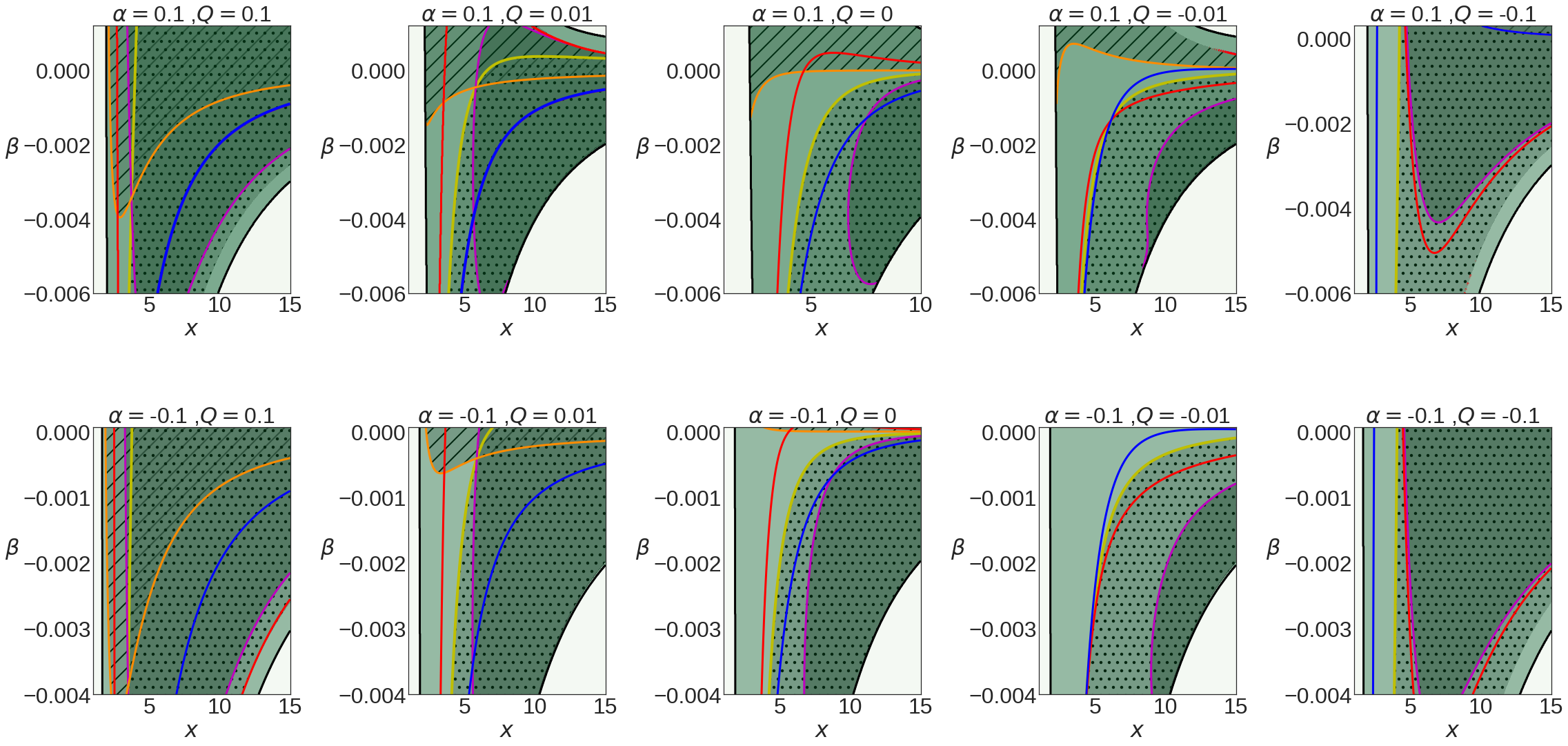}
    \caption{Order of the different epicyclic frequencies in the ($x,\beta$)-plane. Timelike circular orbits can exist in the green light area. The blue curve represents $w_y^2=0$ and the red one $w_x^2=0$. The pink, orange and yellow depicts $w_x^2=\Omega^2$, $w_y^2=\Omega^2$ and $w_x^2=w_y^2$, respectively. The green-gray area shows the region where $w_x^2>\Omega^2$. The hatched line region represents the area where $w_y^2>\Omega^2$. Also, the hatched dotted area is where $w_x^2>w_y^2$. The analysis has done for different values of $Q$. Besides, two different values of the deformation parameter $\alpha$ has been tested.}
    \label{figorder1}
\end{figure*}

Figure \ref{figstab2Zoom} is an extension of Figure \ref{figstab2} only for negative values of $Q$ close to zero. On the three Figures \ref{figstab1}-\ref{figstab2Zoom}, the red line and the blue line correspond to $w_x^2=0$ and $w_y^2=0$, respectively. The dark-green region bounded by those two blue and red lines represents the area where both are positive $w_x^2>0$ and $w_y^2>0$, which is the condition to have stability with respect to vertical and radial oscillations. Note that above the red line, where we have $w_x^2<0$ and $w_y^2>0$, orbits are stable with respect to vertical oscillations but unstable to the radial one. And on the contrary, below the blue line, where $w_x^2>0$ and $w_y^2<0$, the orbits are stable with respect to radial oscillations and unstable with respect to the vertical one. We analyze these Figures by exploring the existence of the timelike circular orbit for different parameters. In fact, the region of existence is clearly affected by the three parameters. In the $(x,\beta)-$ plane in Figure \ref{figorder1}, we see that switching $\alpha$ from negative to positive values tends to shrink the light-green region, on the contrary to an increase in $\beta$. The effect of the magnetic parameter is not monotonic; namely, by decreasing $Q$ from positive to zero, the region is reduced in the vertical direction, but when $Q$ becomes negative and still we continue to decrease it, the region extends again.

Let's focus now on the stability region (dark-green area). As for the existing area, all three parameters also have influence on the region of stability but with different strengths. We can note that switching $\beta$ from negative to positive values makes the area of stability shrink (see Figure \ref{figstab2}). The opposite effect is visible when switching $\alpha$ from negative to positive (see Figure \ref{figorder1}). However, the region of stability hardly changes with the parameter $Q$. It starts to shrink when $Q$ decreases and continues by being pushed away from the central mass and then completely disappear for the negative values. This due to the fact that, on the $(x,\beta)-$ plane, the blue curve comes up, and the red one comes down, so the dark-green region shrinks, and at the end does not exist anymore. For larger negative values, a branch from above comes back in the physical range but keeps staying below the blue curve. On the $(x,\alpha)-$ plane, this is due to the fact that the red curve goes up and then goes out from the physical range. This effect can be seen by combining Figures \ref{figstab2} and \ref{figstab2Zoom}. This situation is mainly like what is happening for negatives values of $Q$. A small area remains for values of the magnetic parameter close to zero $Q=-0.01$. In this case, it means there is no stable orbit in any direction in the chosen physical range.

\begin{figure*}
    \centering
    \includegraphics[width=\hsize]{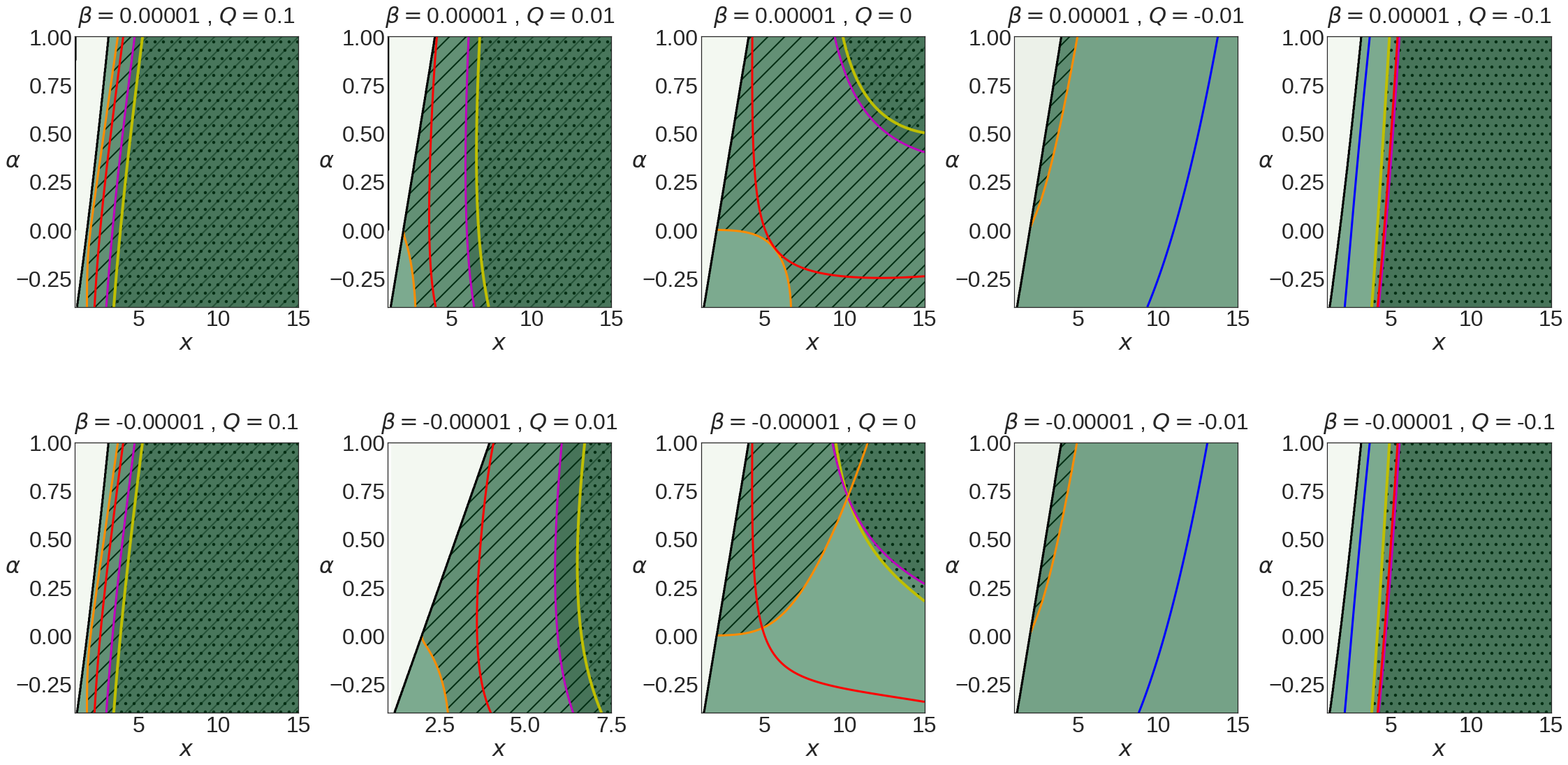}
    \caption{Order of the different epicyclic frequencies in the ($x,\alpha$)-plane. Timelike circular orbits exist in the green light area. The blue curve represents $w_y^2=0$ and the red one $w_x^2=0$. The pink, orange and yellow depicts $w_x^2=\Omega^2$, $w_y^2=\Omega^2$ and $w_x^2=w_y^2$, respectively. The green-gray area shows the region where $w_x^2>\Omega^2$. The hatched line region represents the area where $w_y^2>\Omega^2$. Finally, the hatched dotted area is where $w_x^2>w_y^2$. The analysis has done with the focus on the different signs of $Q$. In addition, two different values of the deformation parameter $\beta$ has been tested.}
    \label{figorder2}
\end{figure*}

In Figures \ref{figorder1} and \ref{figorder2}, different relations between the frequencies have been plotted. Similar to Figures \ref{figstab1} and \ref{figstab2} the red and the blue curves are representing $w_x^2=0$ and $w_y^2=0$, respectively. We add to this picture more information that helps to analyze the order of the different epicyclic frequencies $w_x^2$, $w_y^2$ and $\Omega^2$ easier. The orange line shows $w_y^2=\Omega^2$, which leads to the hatched line region $w_y^2>\Omega^2$. Also, $w_x^2=\Omega^2$ corresponds to the pink line related to the dark-green area where $w_x^2>\Omega^2$. Furthermore, $w_x^2=w_y^2$ is shown by the yellow line. The corresponding region $w_x^2>w_y^2$ is presented in the hatched dotted line. By analyzing both Figures together, we can order the frequencies as a function of the magnetic parameter $Q$. We see that for $Q\geq 0$, the order's behavior will depend on two crossing points. One when the orange line crossed the red line and another when the orange, pink and yellow all are crossing.

For small values of $\beta$, the order is pretty steady (see plots with $Q\geq 0$ on the left in Figure \ref{figorder2}). However, by increasing $\beta$, the crossing points appear, then the behaviour of ordering becomes more complicated (see plots with $Q\geq 0$ on the left in Figure \ref{figorder1}). For instance, using the second plot of the first raw of Figure \ref{figorder1} in the stable region; namely, between the red and blue lines, the frequencies are ordered as follow:

\begin{enumerate}[label=(\roman*)]
    \item Inside the hatched region (above the orange line) where $\beta$ is close to zero, and $\alpha$ is positive or negative:
   \begin{enumerate}[label=(\alph*)]
        \item from the red line to the pink one: $w_x^2<\Omega^2<w_y^2$,
        \item from the pink line to the yellow one: $\Omega^2<w_x^2<w_y^2$,
        \item above the yellow line: $\Omega^2<w_y^2<w_x^2$.
    \end{enumerate}
     \item Outside the hatched region (below the orange line), where $\beta$ has larger negative values, the order of the frequencies is different:
     
    \begin{enumerate}[label=(\alph*)]
        \item from the red line to the yellow one: $w_x^2<w_y^2<\Omega^2$,
        \item from the yellow line to the pink one: $w_y^2<w_x^2<\Omega^2$,
        \item and above the pink line: $w_y^2<\Omega^2<w_x^2$.
    \end{enumerate}
\end{enumerate}
About negative values of $Q$, the order's distribution is different; however, the analysis is simple because the region of stability is strongly reduced. In this Figure, the only stable region appears for the positive value of $\alpha$ and the case $Q=-0.01$ (the fourth plot of the top line in Figure \ref{figorder2}). In this small stable region (dark-green region in the corresponding plot in Figure \ref{figstab2}) the order is as follow:
\begin{enumerate}[label=(\roman*)]
       \item from the red to the yellow line (small extended vertically region): $w_x^2<w_y^2<\Omega^2$,
    \item from the yellow line to the blue one: $w_y^2<w_x^2<\Omega^2$.
\end{enumerate}
To sum up, increasing $Q$ makes the region of specific order larger and creates new regions where one can find different orders for the frequencies.

\section{QPO Models} \label{models}

A class of models, so-called orbital models, assume a relation between the QPO frequencies and the frequencies related to motion of accreted matter orbiting in the vicinity of a compact object. These models focus on hot-spot or disc-oscillation to describe QPOs. However, the existing QPO models are incomplete tasks. Since the oscillations predicted by these models are usually not seen in the same way in the MHD simulations. Furthermore, none of these models has a good match, especially with the full QPOs amplitudes and the visibility on the source spectral data, in the LMXBs. 

Among these various models we consider the group of QPO models considered in \citep{2011A&A...531A..59T} and examine them in this set-up. For a detailed discussion on these models see also \citep{2005A&A...436....1T,2013A&A...552A..10S,2017A&A...607A..69K,2017EPJC...77..860K}. In this section, we begin by explaining the main features of each model briefly. As we see in the following, their prescriptions rely on the formulae for epicyclic frequencies of a particle motion \citep{2011A&A...531A..59T} which are shared the primary motivation with some cases for example models based on the dynamic of fluid \citep{2020A&A...643A..31K}. In fact, in a given model, the predicted QPOs can be expressed in the test particle
motion with reasonable accuracy even for the consideration of discoseismic modes. In the folowing, the two kinematic models are RP and TD, and the resonant models are WD, Ep, Kp, RP1 and RP2. In fact, they consider different possibilities in the combination of disc-oscillation modes. In brief, in the concept of resonance hypothesis \citep{2001A&A...374L..19A}, the two modes in resonance should have eigenfrequencies equal to the radial epicyclic frequency and to the vertical epicyclic frequency or to the Keplerian frequency \citep{2004ragt.meet....1A,2005A&A...436....1T,2011A&A...531A..59T}. Besides, models based on the parametric resonance identify the two observed frequencies of $(\nu_{U}, \nu_{L})$ directly with the eigenfrequencies of a resonance. On the other hand, models based on the forced resonance allows to observe combinational frequencies of the modes. Both parametric and forced resonance models provide precise predictions about the values of observed frequencies considering spin and mass of the observed object in particular in the black hole sources.

\subsection{RPM}

The Relativistic Precession Model (RPM) is one of the first attempts to model QPOs, proposed in \citep{PhysRevLett.82.17,2002nmgm.meet..426S}. In RPM the upper frequency is defined as the Keplerian frequency $\nu_U = \Omega$ and the lower frequency is defined as the periastron frequency i.e. $\nu_{p}:= \nu_L = \Omega - \nu_x$. Their correlations are obtained by varying the radius of the associated circular orbit. Within this framework, it is usually assumed that the variable component of the observed X-ray arises from the motion of “hot-spots” or biting inside the accretion disc on a slightly eccentric orbit. Therefore, due to the relativistic effects, the observed radiation is supposed to be periodically modulated. In this model, frequencies predicted are scaled as $1/M$ for a fixed spin value; therefore, the expected frequency ratio is mass independent. As a weakness of this model is the lack of a generic explanation for the observed $3:2$ frequency ratio.

\subsection{TDM}
Another kinematic model is the Tidal Disruption Model (TDM) presented in \citep{2008A&A...487..527C,2009A&A...496..307K}. This follows also very similar approach as the RPM. In this model, the QPOs are assumed as a result of tidal disruption of large accreting inhomogeneities. In other words, when blobs orbiting the central compact object can be stretched by tidal forces forming the ring-section. However, in this case also the frequency ratio is not reliably constrained. In TDM the frequencies are identified with the frequencies of the orbital motion; namely, the upper frequency is defined as $\nu_U = \Omega+\omega_x$ and the lower frequency is defined as $\nu_L = \Omega$.

\subsection{WDM}
The Warped disc Model (WDM) introduced in \citep{2004PASJ...56..905K}, which is related to non-axisymmetric modes in a warped accretion disc. 
In WDM the upper frequency is defined as $\nu_U = 2\Omega-\omega_x$ and the lower frequency is $\nu_L = 2\Omega-2\omega_x$. In more realistic versions of this model, the higher harmonic oscillations are also considered up to the third order, then frequencies like $3\Omega-\omega_x$ possible to consider. On the contrary with two last models, the ratio of the frequencies is crucial for the model. However, this model suffers from the fact that it
considers a somehow exotic disc geometry that causes a doubling of the observed lower QPO frequency.


\subsection{EpM-KpM}
The Epicyclic resonance Model (EpM) \citep{2005A&A...436....1T} is the simplest variant. It is about considering radial and vertical epicyclic oscillations and relates them to the resonance of axisymmetric disc-oscillation modes. The Keplerian resonance Model (KpM) considers a resonance between the orbital Keplerian and the radial epicyclic oscillations. In EpM the upper frequency is defined as $\nu_U = \omega_y$ and the lower frequency is $\nu_L = \omega_x$. In KpM the upper frequency is defined as $\nu_U = \Omega$ and the lower frequency is $\nu_L = \omega_x$.


\subsection{RP1M-RP2M} \label{rr}
The RP1 model by Bursa in 2005 and the RP2 model \citep{2010ApJ...714..748T}, both consider different combinations of non-axisymmetric disc-oscillation modes. In RP1M the upper frequency is defined as the Keplerian frequency $\nu_U = \omega_y$, and the lower frequency is $\nu_L = \Omega-\omega_x$. In RP2M the upper frequency is defined as $\nu_U =2\Omega- \omega_y$ and the lower frequency is $\nu_L = \Omega-\omega_x$. In the case of slow rotation, their outcome  frequencies of oscillation modes are almost coincide with the frequencies predicted by the RPM.

\begin{figure*}
    \centering
    \includegraphics[width=0.75\hsize]{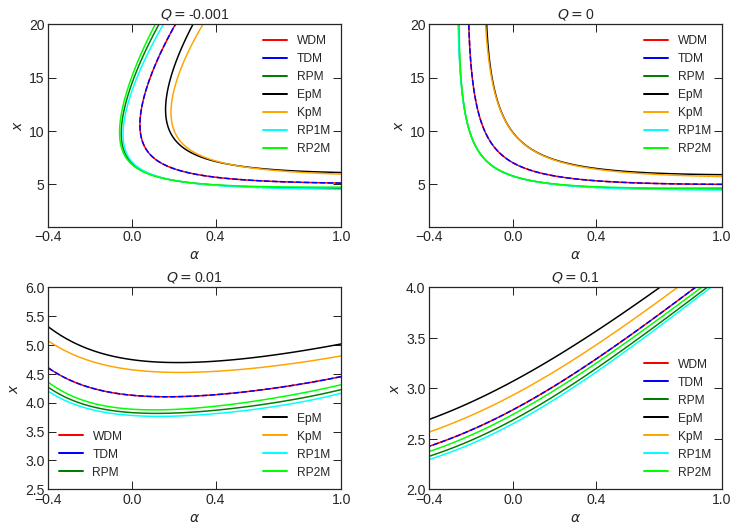}
    \caption{The radii of the $3:2$ frequency ratio for different models. The radius is the function of the $\alpha$ parameter for different values of the magnetic term $Q$. On all the plots, $\beta=0.000001$.}
    \label{rfigModel1}
\end{figure*}
\begin{figure*}
    \centering
    \includegraphics[width=0.75\hsize]{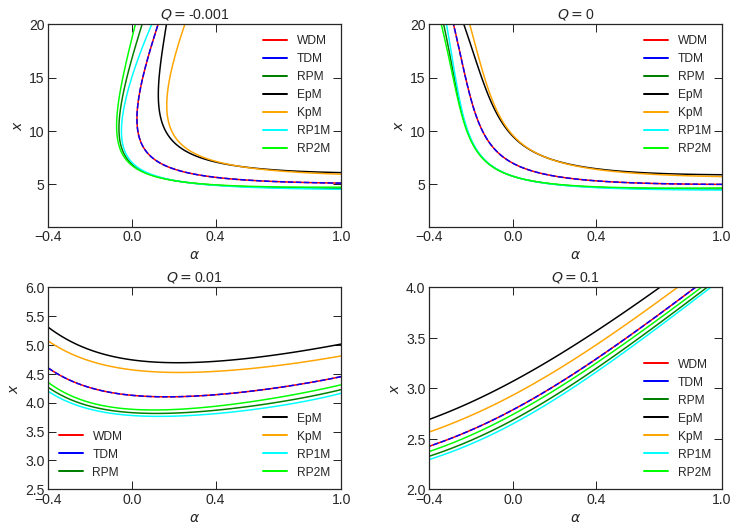}
    \caption{The radii of the 3:2 frequency ratio for different models. The radius is the function of the $\alpha$ parameter for different values of the magnetic term $Q$. On all the plots, $\beta=-0.000005$.}
    \label{rfigModel2}
\end{figure*}

\begin{table}

\caption{\label{Tabelll}Frequency relations corresponding to individual QPO models}
\centering
 \begin{tabular}{c| c| c} 
 \hline
 Model &  $\nu_U$ & $\nu_L$  \\ 
\hline\hline
 RP & $\Omega$ & \quad  $\Omega-\omega_x$\\ 
 \hline
 Kp  & $\Omega$  &  $\omega_x$ \\ 
 \hline
 Ep & $\omega_y$ & $\omega_x$ \\
 \hline
 TD & $\Omega+\omega_x$ & $\Omega$ \\
 \hline
 WD & $2\Omega-\omega_x$ & \quad $2\Omega-2\omega_x$ \\
 \hline
 RP1 & $\omega_y$ & \quad $\Omega-\omega_x$  \\
 \hline
 RP2 & $2\Omega-\omega_y$ & \quad $\Omega-\omega_x$  \\[1ex]
  \hline
 \end{tabular}
\end{table}
The behaviour of these models is illustrated in Figures \ref{rfigModel1} and \ref{rfigModel2}. In these Figures ,the radius of the $3:2$ frequencies for the different models (WD, TD, RP, Ep, Kp, RP1, RP2) is plotted, with respect to the deformation parameter $\alpha$, and for two chosen values of the distortion parameter $\beta$.

This radius depends on all parameters $\alpha$, $\beta$, and $Q$. Nevertheless, on both Figures, Our primary focus is on the impact of the parameter $Q$. As it is seen in Section \ref{sec:PropFreq} where we discussed about the existence and stability of the timelike orbits, only small negative values of $Q$ allow having stability with respect to both vertical and radial oscillations.

From the Figures \ref{rfigModel1} and \ref{rfigModel2} we can see that for negatives values of $Q$, two radii can satisfy the $3:2$ ratio. This situation also happens for all models. Besides, increasing $Q$ tends to move the maximum of the curve to the right. Thus, increasing $Q$ allows choosing larger interval for $\alpha$ where the radius of the $3:2$ ratio exists. Furthermore, increasing the magnetic parameter causes to push this radius closer to the central object. About the different models, we can see that the $3:2$ ratio of the RPM, RP1M, and RP2M appears at the radii similar to each other. As mentioned in the \ref{rr}, this is what we expected for a slow rotating set-up. The same effect is seen for the EpM and KpM. However, models seems to deviate from each other in the case of $Q<0$ in the upper part of the curves. Furthermore, Figure \ref{rfigModel2} shows that in general, the behavior of the negative $\beta$ is very similar to the positive $\beta$ which is discussed earlier. However, slight differences occur on the curves' upper branch for the negative value of $Q$, where in this case, the KpM and EpM diverge from each other. The same is true for RP1M, RP2M, and RPM.

This is worth mentioning that in this deformed and distorted background, with different combinations of parameters $\alpha$, $\beta$ and $Q$, it is possible to have other ratios which can be relevant in other observed data like in other frequencies observed in the Microquasar GRS $1915+105$, see for example \citep{2006astro.ph..7594L}. In addition, the models that associate the variabilities in HF QPQ with discoseismic oscillation modes of the accretion disc, determined that even a weak magnetic field can strongly affect the g-modes  \citep{2009ApJ...690.1386F,2020MNRAS.497..451D}. Therefore, it also seems reasonable to study the  impact of the deformation parameter on these modes. A preliminary investigation reveals that even a slight increase in $\alpha$ changes the corresponding area of the g-mode, manifestly. However, this critical result needs further consideration and can be a future work.

\section{Comparison with the observations}\label{datasec}

\begin{figure}
    \centering
    \includegraphics[width=0.8\hsize]{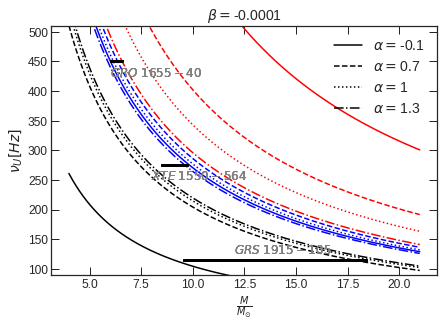}
    \includegraphics[width=0.8\hsize]{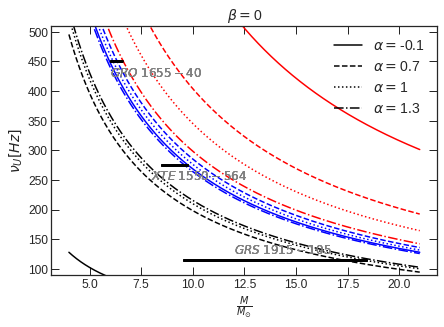}
    \includegraphics[width=0.8\hsize]{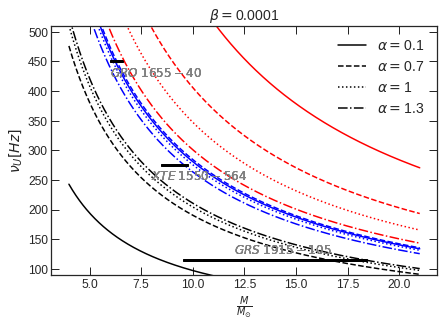}
       \caption{The upper oscillation frequency $\nu_U$ at the resonance radius $3:2$ is presented for various combinations of the studied parameters for the EP model. The magnetic parameter vary with the color's lines. The black lines depict $Q=0$ (unmagnetized case). The blue lines show $Q=0.01$ and the red lines present $Q=0.1$. The upper frequency is compared to the mass-limits obtained from observations of three mentioned microquasars.}
    \label{fig15Data}
\end{figure}

\begin{figure}
    \centering
    \includegraphics[width=0.8\hsize]{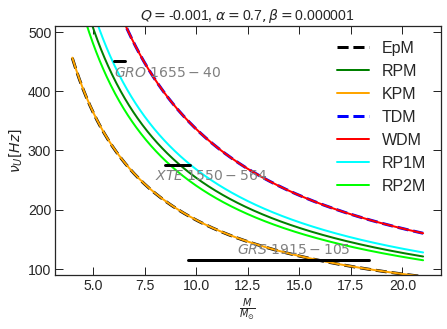}
    \includegraphics[width=0.8\hsize]{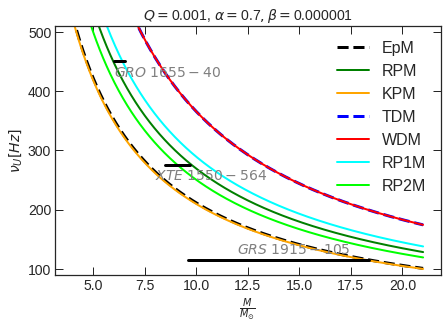}
    \includegraphics[width=0.8\hsize]{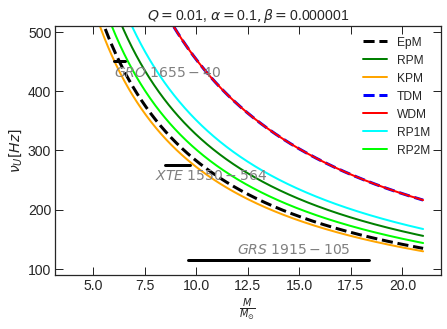}
    \includegraphics[width=0.8\hsize]{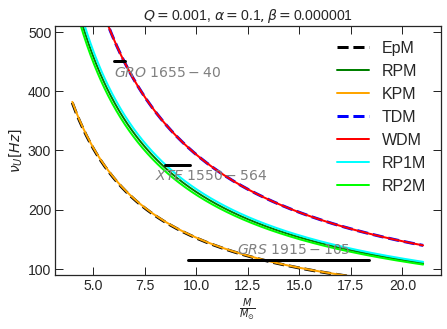}
    \caption{The upper oscillation frequencies for the models presented in the Table \ref{Tabelll} are compared to the mass-limits obtained from observations of three microquasars for various set of parameters.}
    \label{fig16Data}
\end{figure}

The results of fitting the charged particle oscillation frequencies to the observed frequencies of three microquasars XTE $1550-564$, GRS $1915+105$, and GRO $1655-40$ \citep{Remillard2006,Shafee2006} listed in Table \ref{Quasars}, are presented in Figures \ref{fig15Data} and \ref{fig16Data}. Indeed, significant spin is expected in astrophysical black holes at all scales, therefore it is important to develop the model that encodes spin as well, which is in progress. Of course, considering rotation like magnetic field likely to modify the radial profiles of frequencies. Nevertheless, this preliminary results show the fitting can be done even for the fast-rotating microquasar GRS $1915+105$. 

 \begin{table}

\caption{\label{Quasars}Observed HF QPO data for the three micro-quasars, independent of the HF QPO measurement, and based on the spectral continuum fitting.}
\centering
 \begin{tabular}{c c c c} 
 \hline
 Source &  GRO $1655-40$ & XTE $1550-564$ &  GRS $1915+105$ \\ 
 \hline\hline
  $\nu_U$ & $447 - 453$ & $273 - 279$ &  $165 - 171$ \\ 
 $\nu_L$ & $295 - 305$ & $179 - 189$ & $108 - 118$ \\
 $\frac{M}{{M}_{\odot}}$ & $6.03 - 6.57$ & $8.5 - 9.7$ & $9.6 - 18.4$ \\
 $a$ & $0.65 - 0.75$ &  $0.29 - 0.52$ &  $0.98 - 1$ \\[1ex]
 \hline
 \end{tabular}
\end{table}



The observed values of the HF QPO frequencies for this three sources show $\nu_{\mathrm{U}}: \nu_{\mathrm{L}}=3: 2$ ratio \citep{Remillard2006, Shafee2006}. In Figure \ref{fig15Data} we fixed the resonance model EP and try to find the best fitting by exercising different parameters of the model; the deformation $\alpha$, the distortion $\beta$ and the magnetic parameter $Q$.

In Figure \ref{fig15Data} each plot corresponds to different values of the distorted parameter $\beta$, and the line styles present various deformation parameter $\alpha$. The different colors also related to different magnetic parameters. Interestingly, a direct inspection of the various combinations of the parameters shows the crossing with the data lines. While the fitting totally depends on the combination of all parameters, but in general, we can expect to have a better fitting for choosing a larger $\alpha$ in the combination.

In Figure \ref{fig16Data} we have analyzed the different models presented in the last section (see the Table \ref{Tabelll}). Especially, as we have some freedom in choosing the model's parameters, we have decided on ones that minimize computational time and numerical errors and fit the best with data considering Figures \ref{rfigModel1} and \ref{rfigModel2}. 

In fact, we note that although different sets of parameters can fit the data, even for the fast-rotating source as GRS $1915+105$ and GRO $1655-40$; nonetheless, we focus on the fitting to the relatively slowly rotating XTE $1550-564$ source which is more compatible with our set-up. As Figure \ref{fig16Data} suggests, for chosen parameters $Q$, $\alpha$, and $\beta$;  the best fit almost corresponds to RP2M, RPM, and RP1M in low spin cases. KpM and EpM seem to be in favour of the high spin sources. It is worth comparing this result with the behaviour of these models in Kerr spacetime \citep{2011A&A...531A..59T}. In general, it appears that a non-zero magnetic field facilitates the fitting procedure. As a result, we find that WDM and TDM almost have similar behaviour. The same is true for EpM and KpM, while RP1M, RP2M, and RPM may deviate from each other depending on the magnetic field and deformation parameter $\alpha$. These results are almost compatible with the result in Kerr spacetime, where spin plays a similar role to $\alpha$ \citep{2011A&A...531A..59T}. More precisely, by increasing $\alpha$ the deviation between RP's models increases. In this way, studying these three models and the deviations from each other may play an important role to recognize the oblateness of the source from observational data. However, a deeper analysis shows that we can note that almost for $\alpha \in [0.1, 0.9]$ one can have a better fitting to observations. 


Further analysis reveals that the curves take their minimum at different radii depending on the choice of parameters. We see that in all cases, almost the positive and negative values of $\beta$ take their maximum at the same radii; however, further investigation indicates that these radii are smaller for negative values of this parameter $\beta$. Also, the maximums in all curves in the desirable domain depend on the ratio; for example, we see that as this ratio becomes larger, maximum of a curve happens in the smaller radius. This means that the resonance is not monotonic after some distance from the central object, and it depends on the combination of parameters in this set-up. 



\section{Summary and conclusion} \label{sum5}

In this paper, we studied the dynamics of test charged particles in the presence of an asymptotic uniform magnetic field. Further, we have examined different QPO models considered in \citep{2011A&A...531A..59T} in the vicinity of a deformed compact object up to the quadrupole. This space-time is a generalization of the $\rm q$-metric which is static and axisymmetric. It contains two distortion parameter $\beta$ and deformation parameter $\alpha$ which briefly explained in Section \ref{space1}. These two parameters alter the motion and epicyclic frequencies of charged particles moving in this background. However, our main focus was to analyze the influence of magnetic parameters on this set-up. In this manner, the results exhibit a substantial deviation from the nonmagnetic case. It has shown that a regular orbit for some combinations of metric parameters turn to behave chaotically with a magnetic field. Further inspection revealed that magnetic parameter also brings a profound impact on the dynamics and on the radial and vertical frequencies around a stable equatorial orbit.

Additionally, we have shown that the resonant phenomena of the radial and vertical oscillations at their frequency ratio $3:2$, depending on chosen parameters of the model, can be well fitted to the HF QPOs observed in the microquasars GRS $1915+105$, XTE $1550-564$, GRO $1655-40$. In fact, this model may open up a variety of exciting applications in general relativity and astrophysics.

Additionally, it is possible from observational data or other analytic set-ups to assign some restrictions on the parameters in this metric. Moreover, the next step of this work would be to consider models based on the dynamics of fluid in this set-up. This is also possible to explore more about the observational data by considering different inputs in this model, which is the subject of the following work. For example, considering rotation definitely helps to model a more realistic complex system of astronomical objects, or a key direction for future work may be adding the strong magnetic field, which also influences the metric itself.


\appendix 
\appendix 
\section{Epicyclic frequencies in the uniform magnetic field} \label{app1}

The following relations give epicyclic frequencies of particles' circular motion in the background of a distorted, deformed compact object immersed in a uniform magnetic field. The squared vertical frequency is given by


\begin{align} \label{A1}
     w_y^2 = e^{-2\tilde{\gamma}}\left(\frac{x^2-1}{x^2}\right)^{-\alpha(2+\alpha)}\left[ \Omega^2 \frac{x f_1(x,\beta)+S}{S}+(1+f_1(x,\beta)) \Omega \omega_{\text{B}}\right],
\end{align}
And the radial one is given by
 
\begin{align}\label{A2}
     w_x^2 = &\frac{\Omega^2 e^{-2\tilde{\gamma}}(1-1/x^2)^{-\alpha(2+\alpha)}}{x(1-x^2)} \left[g_1(x,\beta,\alpha)\frac{x-S}{S}+g_2(x,\beta,\alpha)\right]\\
     &+ \frac{e^{-2\tilde{\gamma}}(1-1/x^2)^{-\alpha(2+\alpha)}}{x(1-x^2)}
     \left[-\omega^2_Bx(S-x)^2 + \Omega \omega_{\text{B}}g_2(x,\beta,\alpha) \right],\nonumber
\end{align}
where

\begin{align}
&S=1+\alpha+\beta x-\beta x^3,\\
&f_1(x,\beta)=\beta(-1+3x^2),
\end{align}
\begin{align}
&g_1(x,\beta,\alpha) =2 \alpha^{3}+\alpha^{2}\left(6-2 \beta x\left(-1+x^{2}\right)\right)+\\ \nonumber
&2 \alpha\left(2+x\left(x+\beta\left(-1+x^{2}\right)\left(-4+\beta x\left(-1+x^{2}\right)\right)\right)\right)+\\ \nonumber
& x\left(2 x-\beta\left(-1+x^{2}\right)\left(5+x\left(-x+2 \beta \left(-1+x^{2}\right)\left(-3+\beta x\left(-1+x^{2}\right)\right)\right)\right)\right),\nonumber\\
\nonumber
\end{align}

\begin{align}
&g_2(x,\beta,\alpha)= 2 \alpha^{3}-4 x\left(1+\alpha+(-1+\beta) \mathrm{x}-\beta \mathrm{x}^{3}\right)^{2}\\ \nonumber
&-2 \alpha^{2}\left(-3+\mathrm{x}-\beta x+\beta \mathrm{x}^{3}\right)\\
 & +2 \alpha\left(2+\mathrm{x}\left(-2+\beta\left(-1+\mathrm{x}^{2}\right)\left(-4+ \beta x \left(-1+\mathrm{x}^{2}\right)\right)\right)\right)-x\left(-1+x^{2}\right)\nonumber \\
 &\times \left(-1+\beta\left(5+x\left(-4-3 x+2 \beta\left(-1+x^{2}\right)\left(-3+x-\beta x+\beta x^{3}\right)\right)\right)\right).\nonumber
\end{align}

\section*{Acknowledgements}
The authors are grateful to Prof. Stuchlik and Dr. Kolos for a fruitful discussion and the anonymous referee for valuable comments that promoted the work. S.F. thanks the Cluster of Excellence EXC-2123 Quantum Frontiers - 390837967 and the research training group GRK 1620 "Models of Gravity",  founded by the German Research Foundation (DFG). A.T. thanks the research training group GRK 1620 "Models of Gravity", funded by DFG.


\bibliographystyle{unsrt}
\bibliography{bibunm}

\end{document}